\documentclass{SciPost}

\hypersetup{
    colorlinks,
    linkcolor={red!50!black},
    citecolor={blue!50!black},
    urlcolor={blue!80!black}
}

\usepackage[bitstream-charter]{mathdesign}
\DeclareSymbolFont{usualmathcal}{OMS}{cmsy}{m}{n}
\DeclareSymbolFontAlphabet{\mathcal}{usualmathcal}

\fancypagestyle{SPstyle}{
\fancyhf{}
\lhead{\colorbox{scipostblue}{\bf \color{white} ~SciPost Physics Codebases }}
\rhead{{\bf \color{scipostdeepblue} ~Submission }}

\fancyfoot[C]{\textbf{\thepage}}
}

\usepackage{hyperref}
\hypersetup{colorlinks=true, citecolor=blue, urlcolor=blue, linkcolor=blue}
\usepackage{orcidlink}
\usepackage{subcaption}
\usepackage{tikz}
\usetikzlibrary{
  arrows.meta,
  calc,
  fit,
  positioning,
  shapes.geometric
}

\newcommand{\hepdata}{{\sc HepData}}
\newcommand{\rivet}{{\sc Rivet}}
\newcommand{\openai}{{\sc OpenAI}}
\newcommand{\google}{{\sc Google}}
\newcommand{\anthropic}{{\sc Anthropic}}
\newcommand{\gpt}{{\sc Gpt-5.5}}
\newcommand{\claude}{{\sc Claude-Opus-4.6}}
\newcommand{\gemini}{{\sc Gemini-3.5-Flash}}
\newcommand{\geminipro}{{\sc Gemini-3.1-Pro-Preview}}
\newcommand{\agentrivet}{{\sc AgentRivet}}
\newcommand{\amc}{{\sc Madgraph5\_aMC@NLO}}
\newcommand{\pythia}{{\sc Pythia8}}
\newcommand{\analyst}{{\sc Analyst}}
\newcommand{\coder}{{\sc Coder}}
\newcommand{\coderev}{{\sc Code Reviewer}}
\newcommand{\physrev}{{\sc Physics Reviewer}}

\begin{document}
\pagestyle{SPstyle}
\begin{center}{\Large \textbf{\color{scipostdeepblue}{
AgentRivet: an automated system for producing Rivet routines\linebreak from journal publications
}}}\end{center}

\begin{center}\textbf{
Antonio J. Costa\orcidlink{0000-0001-6305-8400}\textsuperscript{1$\star$}
Caterina Doglioni\orcidlink{0000-0002-1509-0390}\textsuperscript{1$\star$}
Christian G{\"u}tschow\orcidlink{0000-0003-0857-794X}\textsuperscript{2$\star$}
Andrew D. Pilkington\orcidlink{0000-0001-8007-0778}\textsuperscript{1$\star$}
Sukanya Sinha\orcidlink{0000-0002-2438-3785}\textsuperscript{1$\star$}
}\end{center}

\begin{center}
{\bf 1} Department of Physics \& Astronomy, University of Manchester, Manchester M13 9PL, United Kingdom
\\
{\bf 2} Centre for Advanced Research Computing, University College London, London WC1E 6BT, United Kingdom
\\[\baselineskip]
$\star$ \href{mailto:agent-rivet-support@cern.ch}{\small agent-rivet-support@cern.ch}
\end{center}

\section*{\color{scipostdeepblue}{Abstract}}
\textbf{\boldmath{%
Particle physics collider experiments provide \rivet{} routines as part of the analysis preservation strategy for model-independent measurements. \rivet{} is a C++ toolkit that allows  new theoretical models to be compared to the measurements, thus aiding the development and tuning of Monte Carlo event generators as well as searches for physics beyond the Standard Model. However, analysis coverage is known to be incomplete, with only 39\% of  measurements having documented and publicly available \rivet{} routines. In this article, we design and implement an automated workflow based on Large Language Models with the goal of providing the missing routines. This multi-step workflow, referred to as \agentrivet{}, extracts the physics analysis information from published papers and writes the missing \textsc{Rivet} routines, with intermediate code- and physics- reviews as part of an autonomous quality control. We report the results obtained using commercial Large Language Models, provided by \openai, \anthropic, and \google, for two 
recent measurements from the ATLAS and CMS experiments. We find that \agentrivet{} produces competent \rivet{} routines with few syntax errors. The physics fidelity of the routines is reasonable and follows the explanations given in the relevant publications.  Nevertheless, physics-implementation issues do arise and are investigated using the artefacts produced by \agentrivet{}. The majority of physics implementation issues arise from subtle-but-ambiguous definitions in the given publication, although some models struggle to implement complex observables even when clear definitions are given.
}}

\vspace{\baselineskip}




\section{Introduction}
\label{sec:intro}

The Standard Model of Particle Physics (SM) describes the fundamental particles that exist in nature and the interactions between them. The predictions of the SM have been extensively verified and tested for over five decades using data collected by experiments at LEP, HERA, the Tevatron, and the Large Hadron Collider (LHC) via an extensive programme of precision measurements, which correct the data for the effects of detector inefficiency and resolution. These measurements, often referred to as model-independent measurements, provide a resource based on collider data that lasts well beyond the lifetime of the collider itself.

The agreed standard for LHC experiments is to provide measurements of observables in \hepdata{} format~\cite{Maguire:2017ypu} and with an associated \rivet{} routine ~\cite{10.21468/SciPostPhysCodeb.36}. \hepdata{} is an open access repository for particle physics experimental data, with thousands of measurements recorded over the last four decades. \rivet{} is a C++ framework for specifying the exact fiducial definitions and observables used in a given measurement within a \textit{routine}. Collectively, the use of \hepdata{} and \rivet{} allow for any theoretical model to be compared to existing data, facilitating both the improvement of Monte Carlo event generators that simulate collision events as well as the search for physics beyond the SM~\cite{10.21468/SciPostPhysCore.4.2.013}. The use of \hepdata{} and \rivet{} is a key component of LHC analysis preservation strategies~\cite{LHCReinterpretationForum:2025zgq}.

Despite the clear benefits that arise when \rivet{} routines are provided, analysis coverage is currently far from complete. For LHC measurements, coverage ranges from 49\% at the ATLAS experiment to 16\% at the ALICE experiment. The Tevatron and HERA experiments have provided \rivet{} routines for about 10\% of their publications, a feature that can be explained by the fact that the experiments were operational before \rivet{} became a de-facto standard for the field. Overall, just 39\% of measurements have \rivet{} routines provided and 230 missing routines have been designated as high priority due to the unique information contained in the associated measurements~\cite{rivet_coverage}. 
The limited coverage may partly reflect a perception amongst analysers that the effort invested into analysis preservation is not yet sufficiently recognised or rewarded within the community. Although recommendations and policy changes are beginning to emerge to encourage preservation activities~\cite{10.21468/SciPostPhys.9.2.022,Campana:2025jmz,DPHEP:2025prk}, additional efforts are needed to streamline the production of preservation artefacts and lower the barriers to their widespread adoption.

A potential solution to this problem is to exploit the generative artificial intelligence (AI) capabilities of Large Language Models (LLMs). LLMs are transformer models~\cite{DBLP:journals/corr/VaswaniSPUJGKP17} designed for natural language processing tasks, with billions of parameters trained on petabytes of text to determine the probability of specific word sequences. Fully trained LLMs can demonstrate emergent properties~\cite{DBLP:journals/tmlr/WeiTBRZBYBZMCHVLDF22,schaeffer2023emergentabilitieslargelanguage} and have been shown to be particularly adept at summarising documents and generating code~\cite{goyal2023newssummarizationevaluationera,DBLP:journals/corr/abs-2107-03374}. The orchestration of LLMs allows the development of so-called AI agents, i.e. workflows that are capable of making autonomous decisions to complete tasks with minimal or no human supervision~\cite{Wang_2024,Faroughy:2026dkj}.

In this article, we present \agentrivet{}, a Python-based AI workflow that accesses journal publications, extracts the relevant physics information, and provides a \rivet{} routine for that analysis. An intermediate review loop assesses the quality of the routine, both in terms of coding and physics implementation. \agentrivet{} can therefore provide a solution for missing \rivet{} routines, but only if the physics accuracy and code quality are sufficiently high. 
We assess the performance of \agentrivet{} using commercially available LLMs provided by \openai, \anthropic, and \google{}.
In addition to potentially solving the \rivet{} routine coverage, we note that this article adds to the growing body of literature that documents the performance and capabilities of AI agents~\cite{Diefenbacher:2025zzn,Moreno:2026jfc,Badea:2026agentic,He:2026drsai,Menzo:2025heptapod,Hill:2026grace,GendreauDistler:2025llmhep,Plehn:2026madagents,Faroughy:2026dkj}.

The article is structured as follows: In Sections~\ref{sec:design} and \ref{sec:prompts}, we present the design of \agentrivet{} along with details of the prompt tuning for the LLMs. Section~\ref{sec:methods} defines the control publications and associated Monte Carlo (MC) event generator samples that are used to assess the performance of \agentrivet{}. The results obtained with \agentrivet{} using the commercially available LLMs are presented and discussed in Sections~\ref{sec:results} and \ref{sec:discussion}. We summarise our findings in Section~\ref{sec:summary}.

\section{Code design and LLM orchestration}
\label{sec:design}

The \agentrivet{} software framework was designed around the principle of modular, provider-agnostic orchestration of large language models (LLMs) for scientific analysis reinterpretation and code synthesis. 
The LLM orchestration is not implemented in a commercial or general-purpose open-source agentic programming framework. Instead, \agentrivet{} implements a lightweight orchestration layer designed to provide direct control over agent interactions while preserving intermediate LLM inputs, outputs, and decisions for inspection and explainability. This allows the same behaviour and contribution of individual agents to be studied explicitly, while permitting the same workflow to operate across multiple LLM providers through a stable internal interface for agent coordination, shared memory, and structured outputs.


At the core of the framework is a lightweight \textsc{Agent} abstraction consisting of a backend LLM interface, a system prompt defining the role and behavioural constraints of the agent, optional tool definitions, and an optional structured output schema. Each agent exposes a common \texttt{run()} interface accepting a user prompt and returning either free-form text or a validated structured object. This allows heterogeneous agents with different responsibilities to be composed into larger deterministic workflows. The workflow described in this paper does not use LLM-native tool calling, in which the model itself requests tool invocations during generation. Instead, operations such as information extraction are performed through deterministic calls orchestrated directly by the workflow, avoiding additional LLM inference steps where agent-driven tool selection offers no clear benefit.

The framework currently employs several specialised agents. An \analyst{} extracts structured analysis information from publications and auxiliary resources, including fiducial phase-space definitions, object constructions, event-selection requirements, and histogram specifications. A \coder{} subsequently generates Rivet-compatible C++ analysis implementations based on this structured representation. The review stage is decomposed into two specialised agents in order to separate software-engineering and physics-validation concerns. A \coderev{} evaluates the generated implementation for potential C++ and Rivet-specific issues, while a \physrev{} compares the generated implementation against the extracted analysis specification and identifies inconsistencies in object definitions, fiducial selections, cuts, and observables. This decomposition reduces the scope and cognitive complexity of each model invocation by restricting each reviewer agent to well-defined tasks.

Inter-agent communication is mediated through a shared state object representing the evolving analysis context. This memory layer stores both structured objects and intermediate artefacts, including downloaded publication text, extracted analysis metadata, generated source code, review comments, and execution diagnostics. The shared-memory abstraction decouples individual agents from one another, allowing each stage to operate only on the subset of information relevant to its task. The memory object can additionally be serialised to disk, permitting expensive LLM-derived intermediate products to be cached and reused across runs in a provider-independent and reproducible manner. This substantially reduces redundant model invocations during iterative development and debugging.

The orchestration layer interfaces with concrete LLM providers through a minimal backend interface. Backend implementations were developed for multiple providers including \openai, \google, and \anthropic{} models, while additionally allowing for extension to locally-hosted or open-source models. Each backend implements a common \texttt{generate()} method operating on a provider-independent message representation. The orchestration layer therefore remains independent of provider-specific SDKs, message formats, or tool-calling conventions. Backend-specific functionality, such as schema-constrained generation or tool execution protocols, is encapsulated within provider adapters. This design permits users to select providers dynamically at runtime while preserving identical orchestration logic across providers.

A key requirement of the workflow is the reliable extraction of structured information throughout the pipeline. To support this, the framework employs structured outputs defined using \textsc{Pydantic} models that store data and annotated fields within a given schema. These models encode the expected schema of extracted analysis information, including beam energies, fiducial definitions, object constructions, event-selection criteria, and histogram specifications, and structured review outputs. LLM responses are validated against these schemas immediately after generation, enforcing consistent downstream data structures and allowing malformed or incomplete outputs to be detected before propagating to later stages of the pipeline. 

A feature of the orchestration framework is the implementation of an iterative review loop between the code-generation and review stages. After an initial code draft is produced by the \coder{}, both the \coderev{} and \physrev{} evaluate the implementation from complementary perspectives and produce structured review objects describing issues and suggested corrections. The resulting review feedback is subsequently incorporated into the next iteration of the \coder{} prompt, allowing the implementation to be progressively refined. The loop continues until either the reviewer signals approval, no major issues remain, or a configurable iteration limit is reached. This iterative refinement mechanism was inspired by conventional software-engineering review workflows and aims both to improve technical robustness of the generated Rivet implementation and to maximise consistency with the original physics analysis description. We note that the \rivet{} routine is not compiled in  \agentrivet{}, a limitation that we leave for a future version, as discussed in Section~\ref{sec:discussion}.

The structured outputs generated at each stage of the workflow are additionally preserved as serialisable \textit{artefacts} within the shared state. This creates an auditable record of the complete analysis-generation process, allowing users to inspect the inputs and outputs of individual agents and to identify the origin of discrepancies in the final implementation. Rather than treating the LLM pipeline as a monolithic black box, the workflow exposes intermediate representations such as extracted analysis metadata, generated code drafts, and review reports. These artefacts provide valuable insight into the decision-making process of the system, facilitate debugging and validation, and improve the transparency and reproducibility of the generated Rivet analyses.

Since commercial LLM APIs may occasionally experience transient failures or rate limits, the framework incorporates explicit retry and validation logic at the backend layer, relying on the last cached step to minimise unnecessary iterations. API exceptions are intercepted and retried with exponential backoff where appropriate, while structured outputs are validated before being propagated to downstream stages. The orchestration layer therefore treats model outputs as untrusted until validated against the expected schema. The framework was additionally designed as a lightweight and extensible Python package with minimal coupling between orchestration logic and provider-specific implementations, permitting straightforward extension to additional providers or specialised agents without modification of the core orchestration layer.

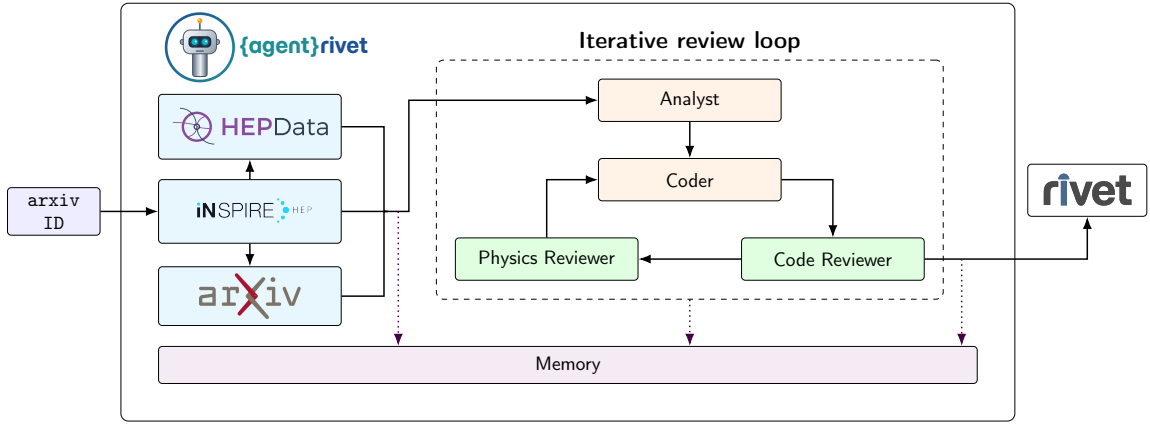
\begin{figure*}[t]
\centering
\scalebox{0.7}{\begin{tikzpicture}[
  font=\sffamily\small,
  >=Latex,
  node distance=12mm and 16mm,
  box/.style={
    draw,
    rounded corners=2pt,
    align=center,
    minimum height=8mm,
    text width=32mm,
    fill=white
  },
  widebox/.style={
    box,
    text width=48mm
  },
  io/.style={
    box,
    fill=blue!7,
    text width=15mm,
  },
  service/.style={
    box,
    fill=cyan!8
  },
  agent/.style={
    box,
    fill=orange!10
  },
  review/.style={
    box,
    fill=green!12
  },
  output/.style={
    box,
    text width=20mm,
  },
  memory/.style={
    box,
    text width=152mm,
    minimum height=7mm,
    fill=violet!8
  },
  group/.style={
    draw,
    rounded corners=4pt,
    dashed,
    inner sep=10pt
  },
  tool/.style={
    draw,
    rounded corners=4pt,
    inner sep=20pt
  },
  flow/.style={->, thick, black},
  data/.style={->, thick, black},
  cache/.style={->, thick, dotted, violet!55!black},
  loop/.style={->, thick, black},
]

\node[memory] (memory) at (42mm,-35mm) {Memory};

\node[io] (cli) at (-55mm,-6mm) {\texttt{arxiv ID}};
\node[service] (scout) at (-18mm,-6mm) {\includegraphics[width=\textwidth]{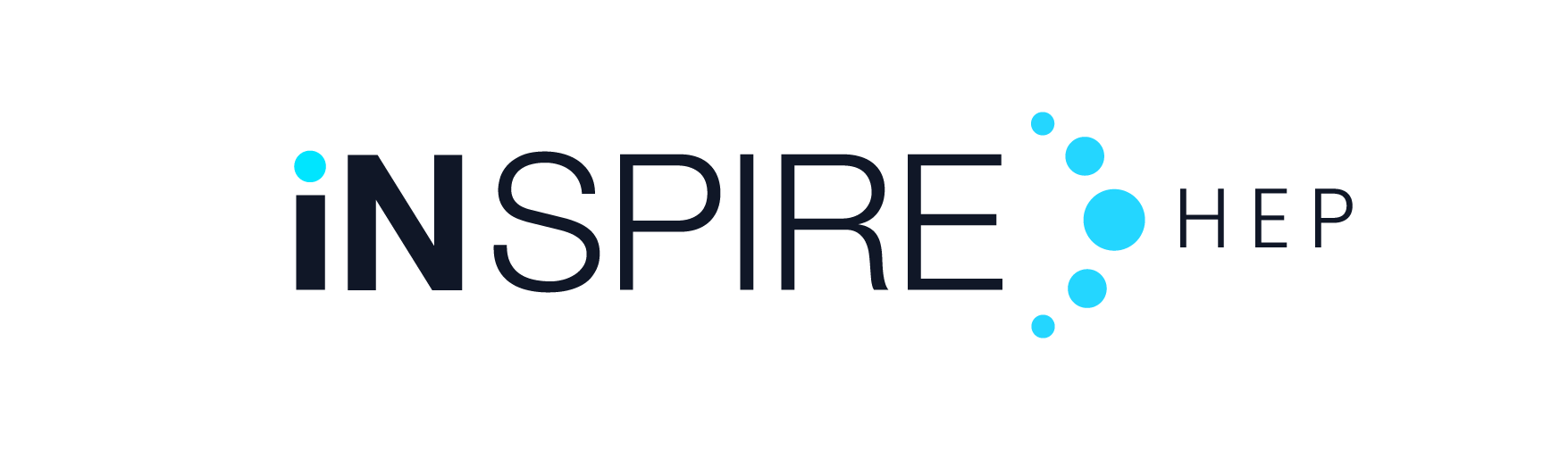}};
\node[service] (hepdata) at (-18mm,10mm) {\includegraphics[width=\textwidth]{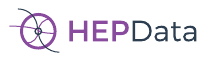}};
\node[service] (pdf) at (-18mm,-22mm) {\includegraphics[width=0.6\textwidth]{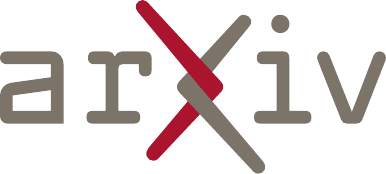}};

\node[agent] (analyst) at (65mm,15mm) {Analyst};
\node[agent] (coder) at (65mm,0mm) {Coder};
\node[review] (physicsReview) at (38mm,-15mm) {Physics Reviewer};
\node[review] (codeReview) at (92mm,-15mm) {Code Reviewer};
\node[group, fit=(analyst)(coder)(codeReview)(physicsReview),
      label={[font=\sffamily\bfseries\large]above:Iterative review loop}] (loopbox) {};

\node (agentrivet-hook) at (-15mm,25mm) {};

\node[tool, fit=(agentrivet-hook)(hepdata)(memory)] (agentrivet) {};


\node[align=center, minimum height=8mm, text width=45mm] (agentrivet) at (-15mm,25mm) {\includegraphics[width=0.9\textwidth]{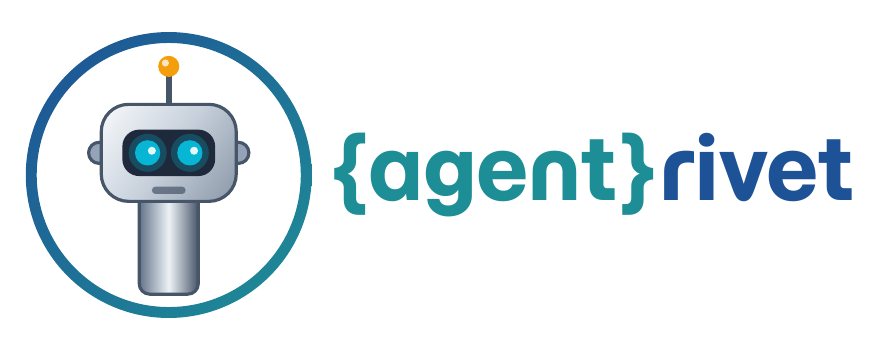}};

\node[output] (cc) at (140mm,-2mm) {\includegraphics[width=0.9\textwidth]{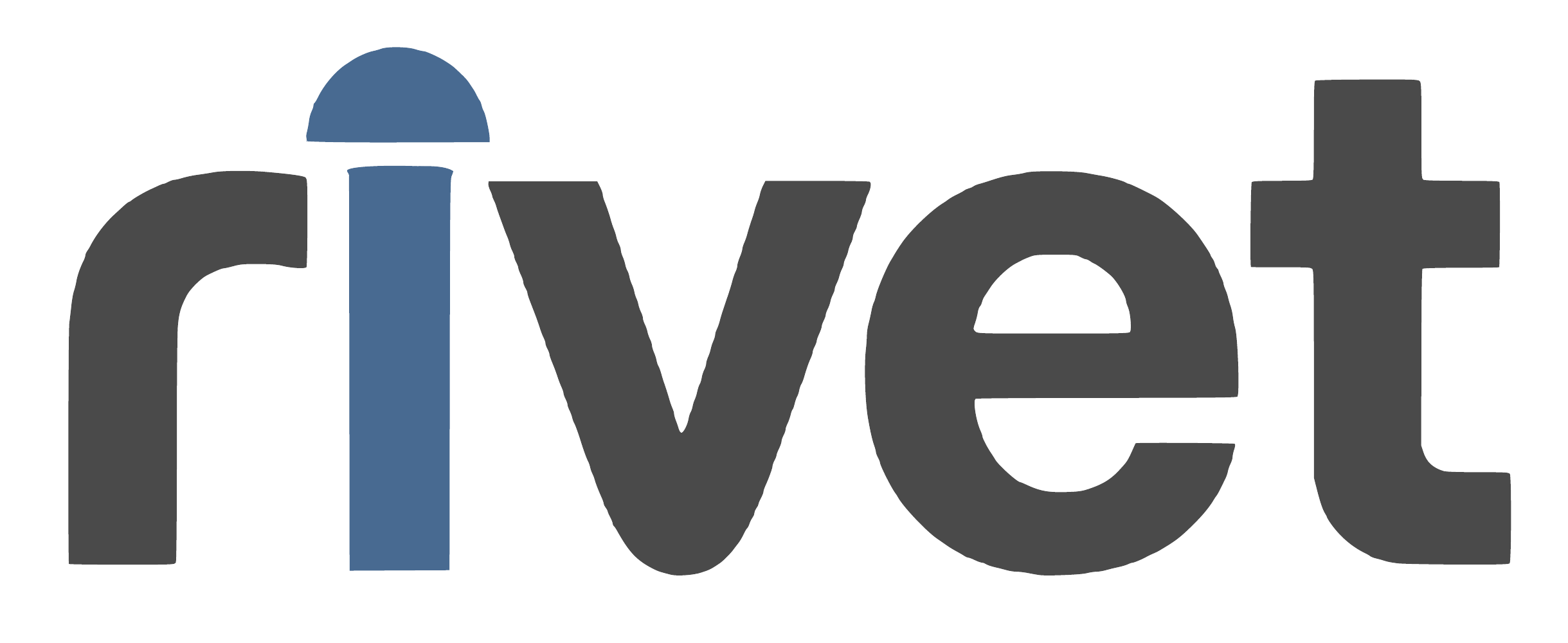}};

\draw[flow] (cli) -- (scout);
\draw[data] (scout) -- (hepdata);
\draw[data] (scout) -- (pdf);

\coordinate (analystInput) at (8mm,-6mm);
\draw[thick, black] (hepdata.east) -- ++(8mm,0) |- (analystInput);
\draw[thick, black] (scout.east) -- (analystInput);
\draw[thick, black] (pdf.east) -- ++(8mm,0) |- (analystInput);
\draw[flow] (analystInput) -- ++(4mm,0) |- (analyst.west);

\draw[flow] (analyst) -- (coder);
\draw[flow] (coder.east) -- ++(2mm,0) -| (codeReview.north);

\draw[loop] (codeReview.east) -- ++(2mm,0) -| (cc.south);
\draw[loop] (codeReview.west) -- (physicsReview.east);
\draw[loop] (physicsReview.north) -- ++(0,2mm) |- (coder.west);

\coordinate (analysisOut) at ($(analyst.east)!0.35!(coder.west)$);
\coordinate (draftOut) at ($(coder.south)!0.5!(codeReview.north)$);
\coordinate (draftOutDrop) at ($(draftOut)+(18mm,0)$);
\coordinate (finalOut) at ($(codeReview.east)+(7mm,0)$);
\coordinate (codeReviewOut) at ($(codeReview.west)!0.45!(physicsReview.east)$);
\coordinate (physicsReviewOut) at ($(physicsReview.north)+(0,9mm)$);
\coordinate (physicsReviewOutDrop) at ($(physicsReviewOut)+(-18mm,0)$);
\coordinate (analystInputWrite) at ($(analystInput)+(2mm,0)$);
\draw[cache] (analystInputWrite) -- (analystInputWrite |- memory.north);
\draw[cache] (loopbox) -- (loopbox |- memory.north);
\draw[cache] (finalOut) -- (finalOut |- memory.north);

\end{tikzpicture}}
\caption{Summary of the \agentrivet{} workflow. The black arrows represent the transfer of information between different steps in the pipeline. The dotted lines represent the storage to Memory.\label{fig:wf}}
\end{figure*}

The workflow of \agentrivet{} is shown in Figure.~\ref{fig:wf} and the code is made publicly available on GitLab~\cite{agent-rivet,agent-rivet-gitlab}, and distributed via the Python package index~\cite{agent-rivet-pypi}.

\section{LLM prompt engineering}
\label{sec:prompts}

The LLM prompts in \agentrivet{} are designed to ensure each agent completes a specific task without replicating the functionality of another LLM. Each prompt has been tuned by inspecting the \agentrivet{} output for two randomly chosen publications~\cite{ATLAS:2025yzd,ATLAS:2025iza} that did not have existing \rivet{} routines. Specifically, points of failure in the workflow were identified during development and testing, and the prompts strengthened to protect against them.

The \analyst{} prompt requires the LLM to provide detailed particle-level information in four main categories: object construction, observable definition, histogram specification and \hepdata{} usage. In each category, explicit rules are given for the information that needs to be extracted. For example, in object reconstruction, photon definitions are requested along with auxiliary information regarding promptness and any isolation requirements. Similar explicit rules are given for jet definitions, lepton definitions and missing transverse momentum. A final rule for \textit{any additional object definitions} provides a fail-safe for non-standard objects. 

The \coder{} prompt specifies the structure of the analysis summary input and the required outputs. Explicit rules are specified for how to use each of the inputs, for example the information in the object construction is to be used to define the \rivet{} projections. The \coder{} is forced to use \rivet4 syntax, with explicit rules for what syntax is forbidden. These syntax rules are found to be necessary to stop the \coder{} mixing \rivet3 with \rivet4 syntax, which are not backwards compatible. Rules are also given on the revision policy, forcing the \coder{} to fix explicit coding errors identified by the \coderev{}, but allowing flexibility on the choice of implementing the suggested fixes provided by the \physrev{}. These choices are discussed further below. 

The \coderev{} is required to only evaluate the code correctness, focussing on incorrect syntax (such as the use of \rivet3 syntax) and likely compile time errors. Explicit rules and examples are given in both cases. The \coderev{} is requested to identify \textit{blockers} that must be acted on by the \coder{}. The review scope of the \physrev{} requires only the physics fidelity to be examined, using the structured analysis summary provided by the \analyst{} as the point of truth. Explicit instructions are given to focus the review on the particle-level object construction, the event selection cuts, and the observable definition. All findings are reported as \textit{advisories} that can be ignored by the \coder{} if the valid \rivet4 syntax is not known. 

\section{Benchmark publications and MC simulations}
\label{sec:methods}

The performance of \agentrivet{} is benchmarked using two recent measurements: inclusive $W\gamma\rightarrow \ell\nu\gamma$ production at ATLAS~\cite{ATLAS:2026xuq} and event shape observables using charged particles inside jets at CMS~\cite{CMS:2026uit}. At the time of writing this article, both publications were publicly available on the arXiv server but no associated \rivet{} routine was provided.\footnote{We note that \rivet{} routines for these very recent measurements are likely forthcoming from the Collaborations.} The two publications provide a stringent test of \agentrivet{} capabilities as discussed below. 

The measurement of inclusive $W\gamma$ production selects events with one lepton, one photon, missing transverse momentum and vetoes on heavy-flavour jets. The particle level is defined using dressed leptons, prompt photons, hadronic jets, and missing transverse momentum, which are standard in precision measurements. Multiple differential cross-sections are measured, including double differential cross-sections, differential cross-sections as a function of angular observables ($\theta_f$, $\phi_f$) defined in a special coordinate system following Lorentz boosts to the centre-of-mass frame, and differential cross-sections as a function of (unfolded) neural network observables. In addition, a binned boost asymmetry observable is defined using the difference between two differential cross-sections divided by the sum of those differential cross-sections. The angular observables provide an important test of the \analyst{}'s ability to identify and communicate the exact definition of complicated observables and the \coder{}'s ability to implement them. The neural network observables are impossible to construct without the neural network model files, thus offering a direct test of LLM hallucinations. The boost asymmetry tests the ability of the coder to calculate the intermediate distributions and then assemble the final result. Furthermore  There was also no public \hepdata{} record and this provides a test of \agentrivet{}'s ability to extract sensible binning from the paper itself. 

The measurement of event shapes using charged particles in jets tests \agentrivet{} in a very different event topology. Events are required to contain at least two jets and the event shape variables are defined using charged particles within those jets. The event shape observables are non-trivial and again offer a stringent test of \agentrivet{}'s ability to construct complicated observables. Finally, a \hepdata{} record exists for this measurement, testing the ability of \agentrivet{} to extract the correct histogram binning information.

Monte Carlo (MC) event generator simulations are produced to test the performance of the \agentrivet{} routines. Inclusive $W\gamma$ events are produced using \amc{} v3.5.15~\cite{Alwall:2014hca} at leading order in perturbative QCD. Parton showering, hadronisation and underlying event activity are added by interfacing to \pythia.316~\cite{Bierlich:2022pfr} to create the particle-level final state. Dijet events are produced using \pythia.316 for testing the event shape analysis.

For each analysis, \agentrivet{} is used to produce \rivet{} routines using the arXiv paper as the point of truth. Three commercial LLMs are investigated, \gpt{} from \openai, \gemini{} from \google{} and \claude{} from \anthropic. Three \rivet{} routines are completed for each LLM setup, to test output consistency as LLMs are inherently non-deterministic. Each \rivet{} routine is compiled using \rivet-4.1.2. In the case that the routine does not compile, the number and nature of the compile errors are noted and the routine is corrected by the authors, making only the minimal changes necessary. Physics fidelity is assessed explicitly for the object reconstruction, fiducial definition and observable construction. Incorrect physics implementation is not corrected by the authors.
The \agentrivet{} routines are used to analyse the Monte Carlo event samples and plots produced for each model and for each run. 

During the journal review of this paper, the official \rivet{} routine for the inclusive $W\gamma$ measurement was made publicly available by the ATLAS Collaboration. The distributions obtained using the official routine are compared to those obtained using the \agentrivet{} routines.


\section{Results} 
\label{sec:results}

\subsection{Reproduction of events shapes measurement}

\begin{figure*}[htbp]
\centering
\begin{subfigure}[b]{0.44\textwidth}
\includegraphics[width=\textwidth]{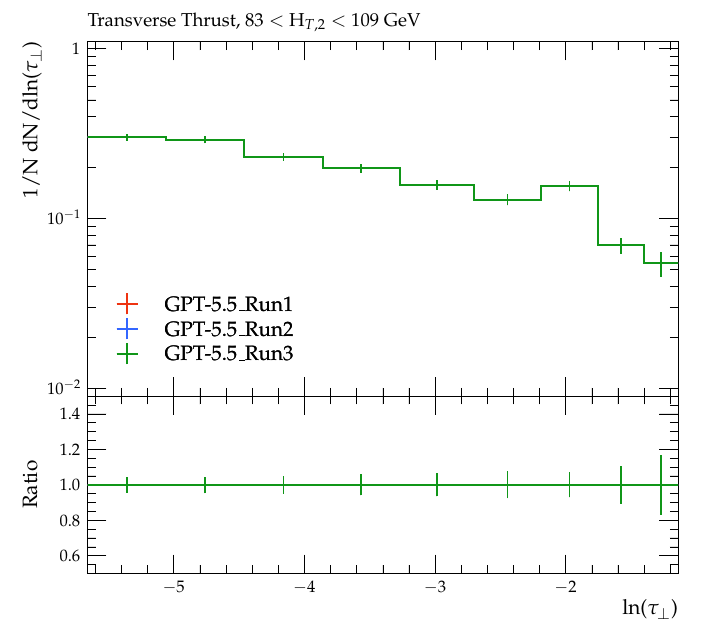}
\caption{}\label{fig:gptthrust}
\end{subfigure}
\qquad
\begin{subfigure}[b]{0.44\textwidth}
\includegraphics[width=\textwidth]{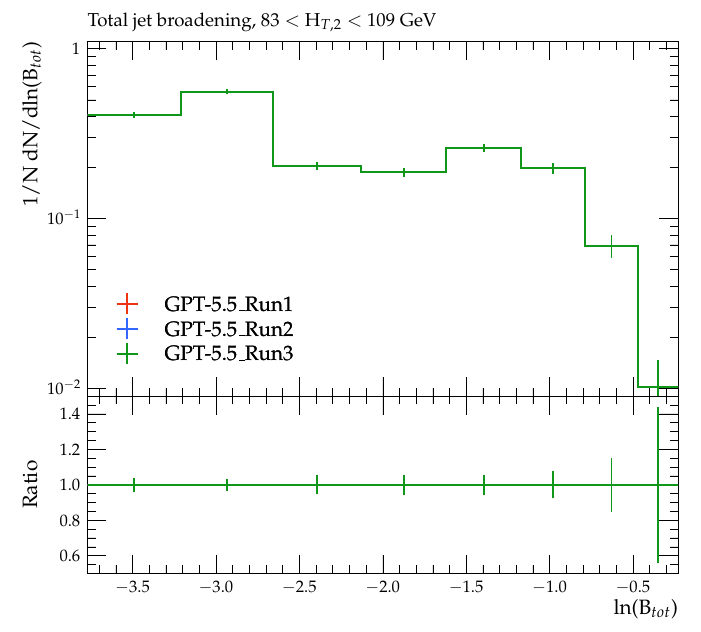}
\caption{}\label{fig:gptbroadening}
\end{subfigure}\\
\begin{subfigure}[b]{0.44\textwidth}
\includegraphics[width=\textwidth]{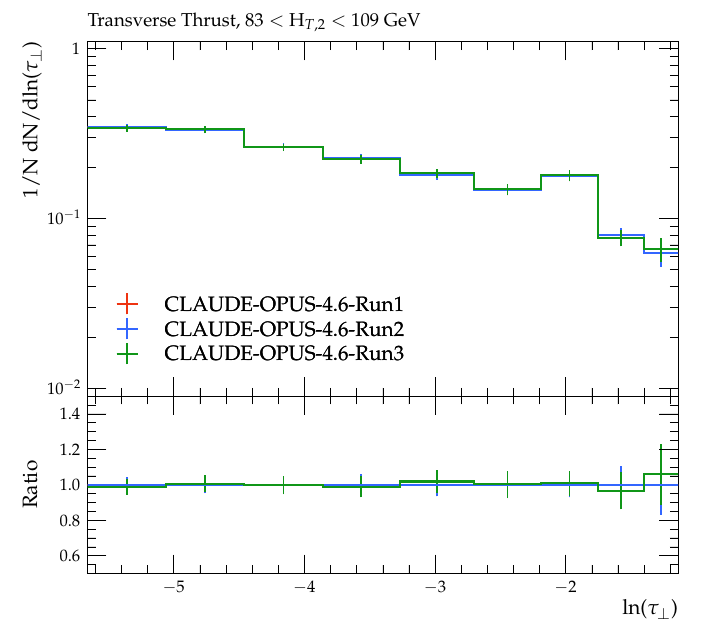}
\caption{}\label{fig:opusthrust}
\end{subfigure}
\qquad
\begin{subfigure}[b]{0.44\textwidth}
\includegraphics[width=\textwidth]{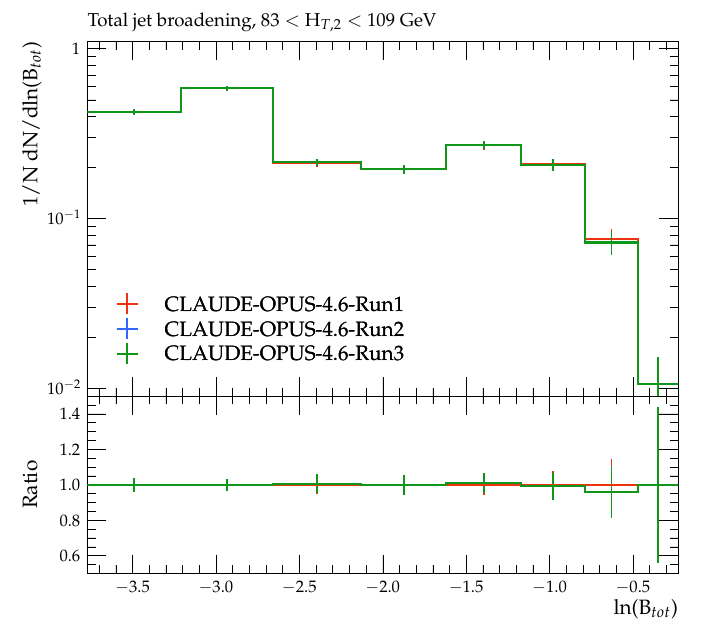}
\caption{}\label{fig:opusbroadening}
\end{subfigure}\\
\begin{subfigure}[b]{0.44\textwidth}
\includegraphics[width=\textwidth]{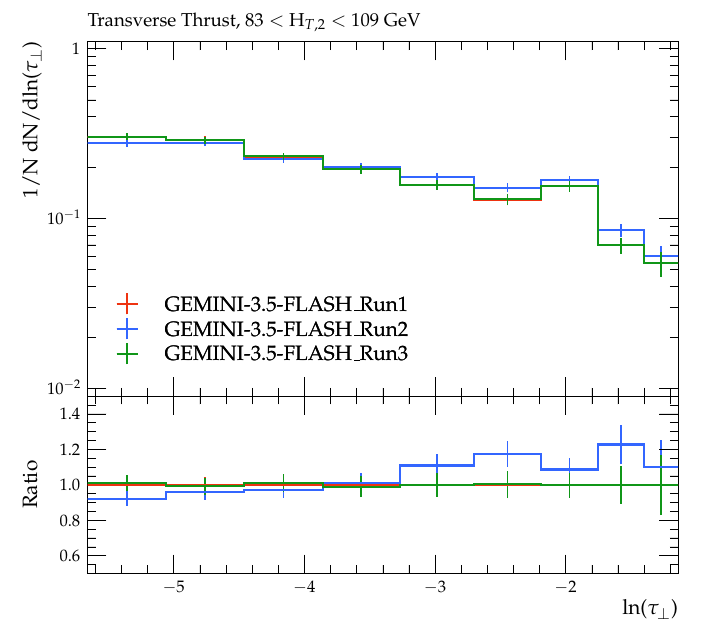}
\caption{}\label{fig:geminithrust}
\end{subfigure}
\qquad
\begin{subfigure}[b]{0.44\textwidth}
\includegraphics[width=\textwidth]{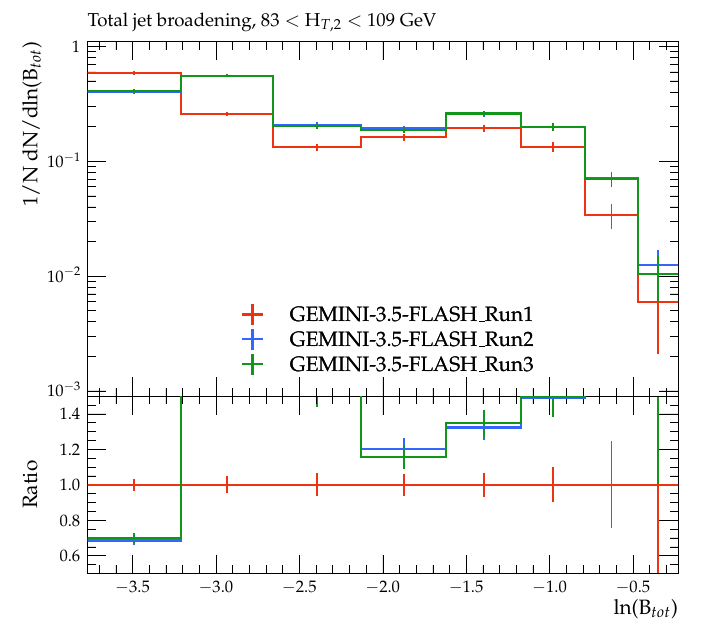}
\caption{}\label{fig:geminibroadening}
\end{subfigure}
\caption{Normalised differential cross-sections as a function of thrust (a,c,e) and broadening (b,d,f) as produced by \agentrivet{} when running on a dijet sample produced using \pythia. Results are obtained with routines produced for \gpt{} (a,b), \claude{} (c,d), and \gemini{} (e,f), with three routines produced for each LLM.\label{fig:evshapes}}
\end{figure*}

\noindent{\textbf{Code quality:}} \agentrivet{} produces routines for the event shapes measurement in jet final states at CMS with minimal issues. All \rivet{} routines produced using \gpt{} are approved by the \coderev{} and \physrev{} without iteration. Routines produced with \gemini{} typically require 2-3 iterations with the \coderev{} to remove deprecated \rivet3 syntax. \claude{} produces final routines but never formally approves them, despite stating that there are zero \textit{blockers} and being explicitly instructed in the prompt to approve routines if no blockers exist. All routines compile without errors when the routine is built (by a human) using \rivet-4.1.2. 
\\
\\ \\ \noindent{\textbf{Object reconstruction and fiducial selection:}} The object reconstruction (jets and tracks) and the fiducial selection are very close to the short description provided in the paper. \gpt{} and \claude{} implement the selection perfectly, whereas \gemini{} implements a veto on final state neutrinos before carrying out the jet finding, despite this not being specified by the paper nor by the \analyst{}; this has a negligible impact on the results.
\\
\\ \noindent{\textbf{Observables:}} The paper measures four observables. Jet mass is correctly calculated by all LLMs in all runs. Thrust and broadening are calculated correctly by \gpt{} and \claude{} for all three runs, but only in two runs of \gemini{}.  
Finally, the third jet resolution parameter is incorrectly calculated by all LLMs, with the calculations varying between different runs of each LLM. This is traced back to the description in the paper providing the formula with an additional (and somewhat ambiguous) statement on a pre-merging step. The \analyst{} is then free to interpret this differently between LLMs and between runs of specific LLMs.
\\
\\ \noindent{\textbf{Histogram binning and normalisation:}} All routines have the correct binning and the histogram identifiers are correctly extracted from the \hepdata{} record. All distributions are normalised by $1/N$ as required, where $N$ is the event yield. However, we found that \gpt{} chooses $N$ to be the event yield within the plot boundaries whereas \claude{} chooses $N$ to be to event yield in the fiducial region (i.e. including underflow and overflow bins). The exact definition of $N$ is not explicitly specified in the publication itself.
\\
\\ \noindent Figure \ref{fig:evshapes} shows results from the \rivet{} routines when applied to the sample of dijet events produced using \pythia{}. Normalised differential cross-sections are presented as a function of thrust and broadening. As discussed above, \gpt{} and \claude{} produce consistent results between runs and agree with each other modulo the $1/N$ normalisation choice. \gemini{} produces an incorrect thrust distribution and an incorrect jet broadening distribution, which in each case is due to an incorrect calculations of the observable.

\subsection{Reproduction of inclusive $W\gamma$ measurement}

\noindent{\textbf{Code quality:}} The production of routines for the inclusive $W\gamma$ measurement at ATLAS is more involved, due to the complexity of the fiducial definition and the wide range of observables measured. Routines require iterations with the \coderev{}, typically passing after one iteration for routines designed by \gpt{} and 2-3 iterations for routines designed by \gemini{}. Routines are never approved by \claude{}. For all LLMs, a small number of errors remain in the routine after the iterative review. For \gpt{}, the remaining code errors relate to deprecated \rivet3 syntax, whereas the errors also include incorrect and hallucinated syntax for \gemini{} and \claude{}.
In one run, \claude{} outputted internal LLM `thinking' narratives on how to tackle the problem at hand, despite the prompt being explicitly designed to avoid this. 
\\
\\ \noindent{\textbf{Object reconstruction and fiducial selection:}} Electrons, muons and photons are correctly reconstructed from prompt particles, defining the charged leptons as `dressed' and removing invisible particles from photon isolation. In the case of \gpt{}, electrons and muons from tau decays are allowed, but this is not explicitly forbidden in the paper description. Jet reconstruction is broadly correct although all routines have slight issues when excluding prompt leptons from the jets: (i) bare charged leptons tend to be excluded instead of dressed leptons, (ii) prompt neutrinos tend to be included in the jet-finding algorithm, a feature that might arise due LLMs interpreting the statement `excluding prompt leptons' in the paper to mean `exclude prompt charged leptons'. 
We noted one instance of \claude{} correctly implementing the exclusion of dressed leptons and one instance of the \physrev{} advising the use of neutrinos in the jets as being likely problematic.

The fiducial selection is implemented correctly in most cases, although a small number of errors are observed whose occurrence depends on both the choice of LLM and the stochastic nature of individual runs. For example, incorrect overlap-removal criteria were generated in one run of \gpt{} and one run of \gemini{}. In addition, \gemini{} almost always applied an incorrect transverse-momentum threshold when implementing the veto on additional dressed leptons. In contrast, \claude{} never misimplemented the lepton veto, while only a single such error was observed for \gpt.
\\
\\ \noindent{\textbf{Observables:}} Observables are correctly reconstructed for all LLMs for all standard  observables, such as transverse momenta, pseudo-rapidity, and invariant mass.  For the angular observables constructed in the centre-of-mass frame of the diboson system, \gpt{} constructs the centre-of-mass frame coordinate system correctly. However, there are two possible solutions for each angular observable due to incomplete information on the neutrino longitudinal momentum. \gpt{} correctly chooses one solution at random in two runs (as specified in the journal publication) but uses both solutions when filling the histogram in one of the runs. \claude{} implements the angular observables correctly for two runs, but applies a critical constraint on the $W$ boson mass to the $\ell\nu\gamma$ system instead of the $\ell\nu$ system in one run. This implementation issue likely arises due to the choice of phrasing in the paper that mentions both the $W\rightarrow \ell\nu$ decay and the $\ell\nu\gamma$ system in the sentence that describes the $W$ mass constraint. 
\gemini{} does not attempt to construct the angular observables at all, a feature that can be traced back to incomplete information provided by the \analyst{}, where the observables are simply stated to be defined `in a special reference frame' despite the exact definition of the reference frame being given in the paper. The \coder{} then does not attempt to construct the angular observables at all. The boost asymmetry observables are also constructed correctly by \gpt{} and \claude{}, but are ignored in two \gemini{} runs and incorrectly implemented in the third. Finally, the LLMs do not attempt to construct the neural-network-based observable as the information is not available.
\\
\\ \noindent{\textbf{Histogram binning and normalisation:}} With no \hepdata{} record, the LLMs have to guess the binning from the available plots in the paper and the binning for each routine is therefore not an exact match to that of the published analysis. 
The standard \rivet{} normalisation procedure is then typically applied to obtain differential cross-sections, although \gemini{} fails to normalise the distributions at all in one run. Interestingly, the double differential distributions are implemented by \gpt{} and \claude{} as consecutive slices of one dimensional distributions, exactly as presented in the paper. This involves mapping the the bins of a two-dimensional distribution to integer values. The standard \rivet{} normalisation procedure then fails because the division by bin width is applied and defaults to the integer bin widths instead of the double differential area. 

\begin{figure*}[htbp]
\centering
\begin{subfigure}[b]{0.44\textwidth}
\includegraphics[width=\textwidth]{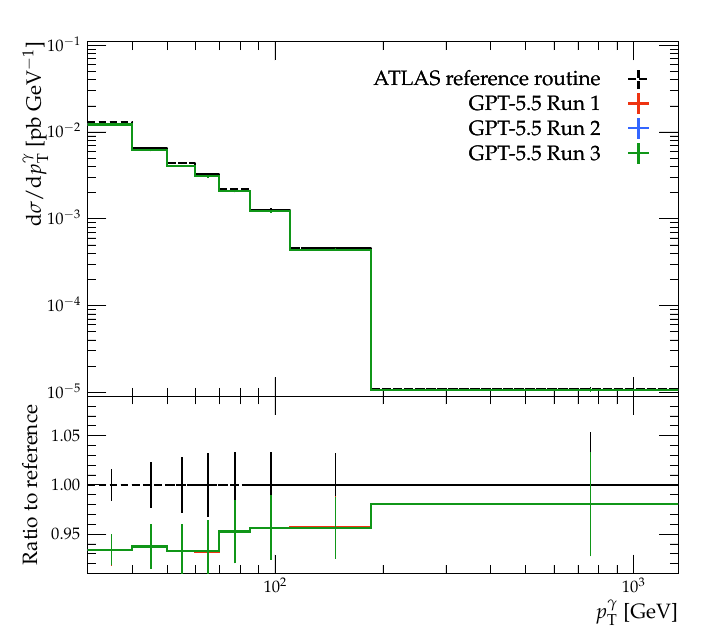}
\caption{}\label{fig:gpt-ptgamma}
\end{subfigure}
\qquad
\begin{subfigure}[b]{0.44\textwidth}
\includegraphics[width=\textwidth]{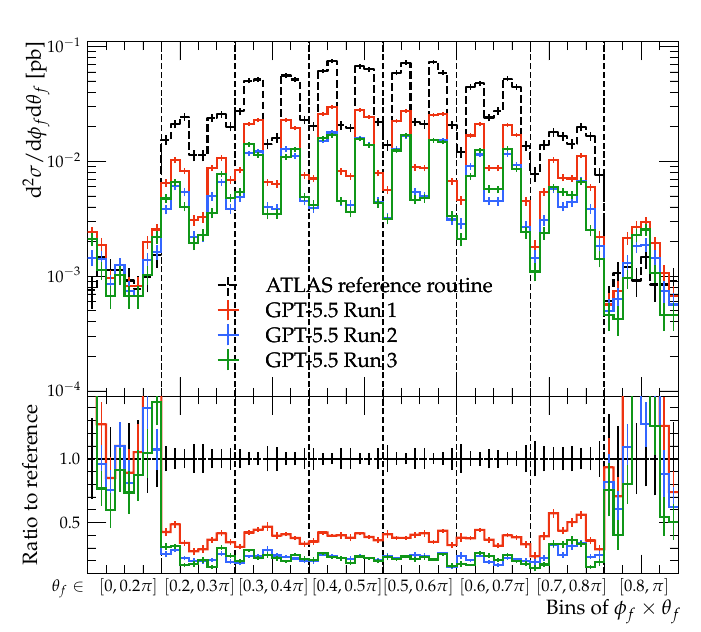}
\caption{}\label{fig:gpt-double}
\end{subfigure}\\
\begin{subfigure}[b]{0.44\textwidth}
\includegraphics[width=\textwidth]{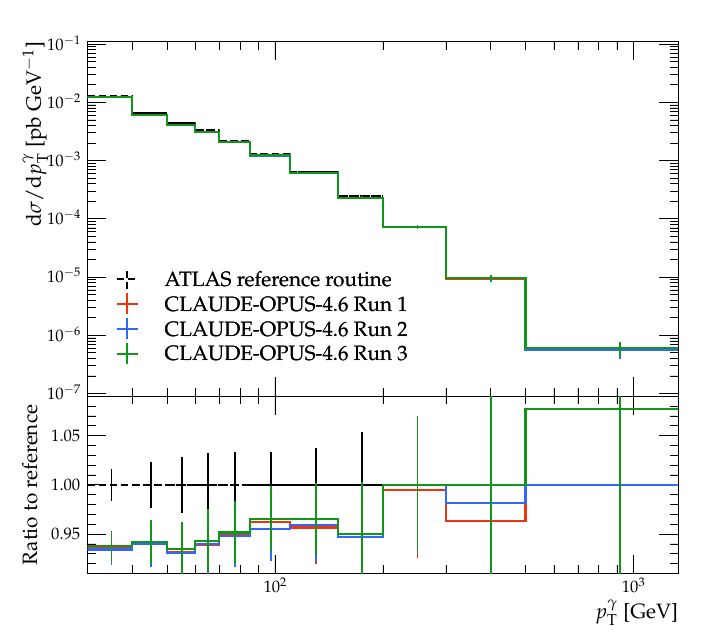}
\caption{}\label{fig:claude-ptgamma}
\end{subfigure}
\qquad
\begin{subfigure}[b]{0.44\textwidth}
\includegraphics[width=\textwidth]{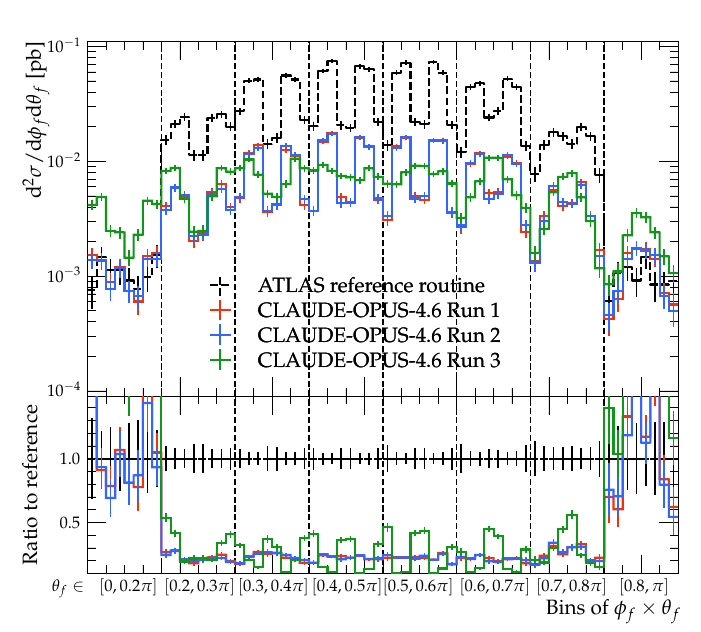}
\caption{}\label{fig:claude-double}
\end{subfigure}\\
\begin{subfigure}[b]{0.44\textwidth}
\includegraphics[width=\textwidth]{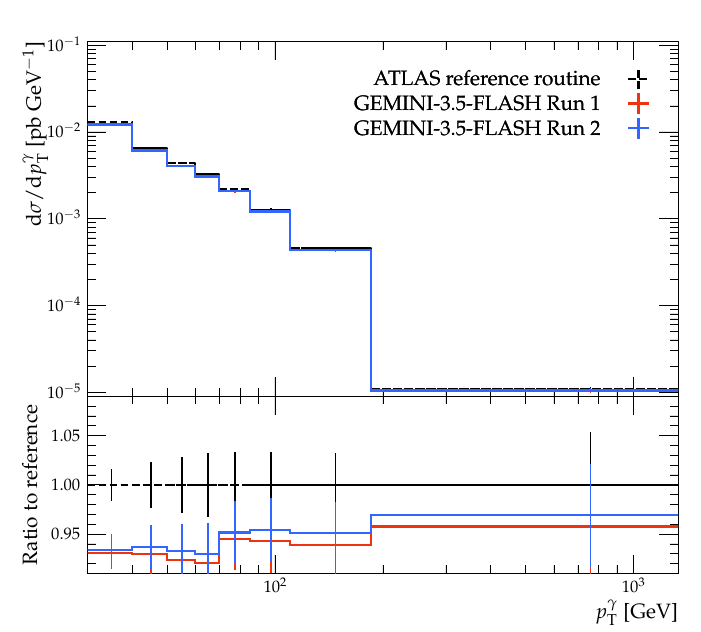}\hspace{0.46\textwidth}
\caption{}\label{fig:gemini-ptgamma}
\end{subfigure}
\caption{Differential cross sections as a function of $p_\mathrm{T}^\gamma$ (a,c,e). Double differential cross sections as a function of $\theta_f$ and $\phi_f$ as measured in the centre-of-mass frame of the $W\gamma$ system (b,d). Results are obtained with routines produced for \gpt{} (a,b), \claude{} (c,d), and \gemini{} (e), with three routines originally produced for each LLM.\label{fig:incwy}}
\end{figure*}

\vspace{0.5cm}
\noindent Figure \ref{fig:incwy} shows results from the \rivet{} routines when applied to the sample of $W\gamma$ events produced using \amc{} and \pythia. The differential cross-section as a function of the photon transverse momentum is shown as well as the double differential cross-section as a function of the angular observables in the centre-of-mass frame ($\theta$ and $\phi$). The results from the \agentrivet{} routines are compared to the official ATLAS \rivet{} routine, which became publicly available during the journal review of this article. For the photon transverse momentum, the \agentrivet{} routines agree with the official ATLAS routine within 5\%, with the small differences likely due to different choices  in object reconstruction and fiducial definition (discussed above). For the  
double differential cross-section as a function of $\theta$ and $\phi$ there are a number of interesting features. The distribution produced by \gpt{} Run-1 is a factor of two larger than the other \gpt{} runs, due to using both neutrino solutions instead of choosing one at random (discussed above). \claude{} Run-3 disagrees with the other two runs due to applying an incorrect $W$ boson mass constraint (discussed above). The distributions produced by \gpt{} Runs 2 and 3 are in reasonable agreement with \claude{} Runs 1 and 2, but a factor of four smaller than the official ATLAS routine. This offset is due to the missing division by bin width for this one-dimensional representation of a two-dimensional distribution (discussed above).

\section{Discussion of AgentRivet performance and features} 
\label{sec:discussion}

A key feature of \agentrivet{} is the permanent storage of the intermediate artefacts that document the analysis summary, source code drafts and review comments. Inspection of these artefacts provides a deeper understanding of the \agentrivet{} output. The structured analysis information output by the \analyst{} is found to be of consistently high quality and provides a strong foundation for the \coder{} to produce a first draft. The review loop is then found to be critical for complex analyses in two ways. First, the \coderev{} uncovers likely compile time errors associated with the incorrect use of \rivet3 syntax and these are usually  corrected on the first iteration. 
The \physrev{} then highlights possible issues, focusing on both missing and incorrect physics implementations. This part of the review is important when calculating the more complex angular observables in the inclusive $W\gamma$ analysis, which require multiple Lorentz boosts and the construction of a special coordinate system. The ability of the  \coder{} to reject suggestions is also important. For example, the \physrev{} continually insists on calculating  neural-network based observables, which is impossible, and the \coder{} correctly rejects this request.

The code quality produced by \agentrivet{} is still not perfect, with some residual \rivet3 syntax making it through the iterative review despite the prompt containing (i) explicit instructions to use \rivet4 syntax, and (ii) examples of deprecated \rivet3 syntax. We find that the specific incorrect usage that is present in the routines is not explicitly listed in the prompt as being deprecated. There are a few possible solutions to this problem. First, the list of deprecated syntax could be expanded in the \coder{} and \coderev{} prompts. Manually adding deprecated syntax when encountered is, however, not a long-term solution and there are more systematic alternatives. First,  compilation feedback could be incorporated into the workflow, allowing compiler diagnostics to be passed back to the \coder{} to inform subsequent revisions. Second, the deprecated syntax could be autonomously updated in the \coder{} prompt, helping to reduce depracated syntax issues in future \agentrivet{} runs. Finally, a fine-tuned open-weights language model (such as Qwen2.5-Coder~\cite{Hui:2024qwen25coder}) could be 
trained specifically on \rivet4 routines, to incorporate domain-specific 
code generation at the expense of requiring a dedicated training pipeline and 
a sufficiently large corpus of high-quality examples. 
This would have the advantage of being a local model and would partially circumvent LLM API access problems, which are discussed next. We leave the implementation and evaluation of these approaches to future work.

Physics fidelity in routines produced by \agentrivet{} is good overall, especially when using \gpt{} and \claude{} backends. However, the quality of the physics implementation is process- and LLM- dependent. Furthermore, different issues can arise within different runs using the same LLM. The majority of these physics-implementation issues can be traced back to non-explicit definitions in the journal publication itself. The choices made by \agentrivet{} in such cases are therefore reasonable. Avoiding such ambiguities is one of the main reasons that LHC collaborations are encouraged to provide Rivet routines in the first place. \gemini{} struggles with the more complex observables measured for inclusive $W\gamma$ production, a feature that is traced back to the \analyst{} providing too little information.

Finally, we comment on issues around API access. \agentrivet{} incorporates an explicit retry and validation logic in case of API being temporarily unavailable. If the API cannot be accessed after a certain number of retries, the artefacts are stored and the run can be resumed later. We find that API-related problems are provider specific and vary with time of day. Originally, we planned to produce results using \geminipro{}, but API access became almost entirely unavailable and runs could not be completed even with multiple manual interventions. The release of \gemini{} in mid-May provided much more stable API access. The \openai{} models were the most accessible and required fewest (if any) manual interventions to complete a given run. Our design choice to provide access to multiple providers and models in \agentrivet{} is therefore a critical feature, as it provides redundancy against unexpected loss of API access for a specific provider. 

At the time of testing \agentrivet{}, the cost of producing a Rivet routine was between USD1.20 and USD2.20, depending on the provider and complexity of the analysis. 

\section{Conclusion}
\label{sec:summary}

In this article, we have presented \agentrivet{}, an autonomous workflow designed to generate Rivet routines directly from particle-physics publications. The workflow combines structured information extraction, code generation, and iterative code- and physics-review stages to produce \rivet4-compatible analysis implementations while preserving a transparent record of intermediate artefacts. 

Using recent measurements from the ATLAS and CMS collaborations as benchmarks, we find that \agentrivet{} is capable of producing \rivet{} routines with very few syntax errors. The physics fidelity is more variable, however, being both process and observable dependent as well as changing between different runs of the same LLM. Most residual physics-implementation issues arise from ambiguities in the original publication, or from particularly complex observable definitions, rather than from fundamental limitations of the workflow itself. The iterative review process is found to be essential for improving both code quality and consistency with the published analysis. Producing routines and results with multiple runs and multiple LLMS  allows problems in specific runs/models to be identified and the routines rejected. However, whilst this would add confidence that a subset of the routines are robust, it would drive up the overall cost of producing them.

The overall performance of \agentrivet{} demonstrates that modern LLMs can successfully extract detailed analysis definitions from scientific publications and translate them into executable scientific software.  Future work will focus on validating the results against the official  routines provided by the experimental collaborations, using dedicated fine-tuned coding models that have been trained on existing \rivet4 routines, and incorporating static-code analysis to identify compilation errors and thereby improve the code review. In addition, we intend to estimate the energy consumption of \agentrivet{} when producing \rivet{} routines. 

\medskip
\noindent {\bf{Acknowledgments}} ---
A.J.C, A.D.P.\ and C.D.\ are supported by the Science and Technologies Facilities Council (STFC) under grant UKRI:2850.
C.G.\ is supported by STFC Enabling AI4HEP under grant UKRI:3900.
This paper is part of a project that has received funding from the European Research Council (ERC) under
the European Union's Horizon 2020 research and innovation programme, (Grant Agreement no.~101002463), supporting C.D.\ and S.S. 

\bibliography{references.bib}
\end{document}